\RequirePackage{ifpdf}
\ifpdf 
\documentclass[pdftex]{sigma}
\else
\documentclass{sigma}
\fi

\def\ba{\begin{array}}
\def\ea{\end{array}}

\def\de{\delta}

\def\l{\langle}
\def\r{\rangle}

\DeclareMathOperator{\wt}{wt}

\numberwithin{equation}{section}

\begin{document}

\allowdisplaybreaks

\renewcommand{\PaperNumber}{091}

\FirstPageHeading

\ShortArticleName{1D Vertex Models Associated with a Class of
Yangian Invariant HS Like Spin Chains}

\ArticleName{One-Dimensional Vertex Models Associated\\ with a Class of
Yangian Invariant Haldane--Shastry\\ Like Spin Chains}

\Author{Bireswar BASU-MALLICK~$^\dag$ and Nilanjan BONDYOPADHAYA $^\ddag$ and Kazuhiro HIKAMI~$^\S$}

\AuthorNameForHeading{B.~Basu-Mallick, N.~Bondyopadhaya and K.~Hikami}

\Address{$^\dag$~Theory Group, Saha Institute of Nuclear Physics,
1/AF Bidhan Nagar, Kolkata 700 064, India}
\EmailD{\href{mailto:bireswar.basumallick@saha.ac.in}{bireswar.basumallick@saha.ac.in}}

\Address{$^\ddag$~Integrated Science Education and Research Centre, Siksha-Bhavana, Visva-Bharati,\\
\hphantom{$^\ddag$}~Santiniketan 731 235, India}
\EmailD{\href{mailto:nilanjan.bondyopadhaya@saha.ac.in}{nilanjan.bondyopadhaya@saha.ac.in}}

\Address{$^\S$~Department of Mathematics, Naruto University of Education, Tokushima 772-8502, Japan}
\EmailD{\href{mailto:KHikami@gmail.com}{KHikami@gmail.com}}

\ArticleDates{Received September 06, 2010, in f\/inal form November 30, 2010;  Published online December 10, 2010}

\Abstract{We def\/ine a class of  $Y(sl_{(m|n)})$  Yangian invariant
Haldane--Shastry (HS) like spin chains, by assuming that their partition
functions
can be written in a particular form in terms of the super Schur polynomials.
Using some properties of the super Schur polynomials,
we show that the partition functions of this class of
spin chains are equivalent to the partition functions of
a class of one-dimensional vertex models with appropriately def\/ined energy
functions.
We also establish a boson-fermion duality relation for the partition functions
of this class of supersymmetric HS like spin chains by using their
correspondence with
one-dimensional vertex models.}

\Keywords{Haldane--Shastry spin chain; vertex model; Yangian quantum group;
              boson-fermion duality relation}

\Classification{81R12; 81R50; 82B10; 82B05}

\section{Introduction}

It is well known that one-dimensional (1D) quantum integrable
 spin chains can be divided into two
classes depending on their range of interaction. One such class consists of
spin chains having only local interactions like nearest
or next-to-nearest neighbor interaction. These spin chains are usually solvable
through coordinate or algebraic Bethe-ansatz technique \cite{Su04,KB93,EF05}.
Isotropic and anisotropic versions
of Heisenberg spin chain, Hubbard model etc. are examples of this type of
spin models with short range interaction. It is worth noting that,
there exists a connection between such quantum spin chains
and vertex models in classical statistical mechanics. More precisely, the
Hamiltonians of 1D
quantum spin models are related to the dif\/ferentiated transfer matrices of the
corresponding 2D vertex models in statistical mechanics \cite{Su70, Ba71}.

Apart from the above mentioned class of spin chains,
there exist another class of 1D quantum integrable spin systems for which
all constituent spins mutually interact with each other through long range
forces.
Haldane--Shastry (HS) spin chain \cite{Ha88,Sh88} and Polychronakos spin chain
(also known as Polychronakos--Frahm spin chain) \cite{Po93,Fr93,Po94}
associated with the $A_N$ type of root system
are well known examples of quantum integrable spin systems with long range
interaction.
This type of spin chains
have attracted a lot of attention in recent years due to their exact
solvability and applicability in a wide range of subjects like fractional
statistics
\cite{Ha91,Ha96, BGS09, BH09},
SUSY Yang--Mills theory and string theory \cite{BD04,ST04,BS05},
Yangian quantum group \cite{HHBP92, BGHP93, Hi95,HBM00, BFLMS10} etc.
The lattice sites of HS spin chain are uniformly distributed on a circle and,
for the case of most general $su(m|n)$ supersymmetric version of this model,
each lattice site is occupied by either one of the $m$ type of bosonic
`spins' or one of the $n$ type of fermionic `spins'.
The Hamiltonian of such $su(m|n)$ HS spin chain is given by \cite{Hal93,BMN06}
\begin{gather}
\mathcal{H}^{(m|n)}_{\rm HS} = \frac{1}{2}  \sum_{1 \le j<k \le N}
\frac{1+\hat{P}_{jk}^{(m|n)}}{\sin^2(\xi_j-\xi_k)}, \label{a1}
\end{gather}
where $\xi_j = j \pi /N$  and $\hat{P}_{jk}^{(m|n)}$ represents the
supersymmetric exchange operator which interchanges the spins
on the $j$-th and $k$-th lattice sites. On the other hand,
the lattice sites of Polychronakos spin chain are non-uniformly distributed on a
line and
the Hamiltonian of the $su(m|n)$ supersymmetric version of this model is given
by
\cite{BMUW99,HBM00}
\begin{gather}
\mathcal{H}^{(m|n)}_{\rm P}  =  \sum_{1 \le j<k \le N}
\frac{1+\hat{P}_{jk}^{(m|n)}}{(x_j-x_k)^2}, \label{a2}
\end{gather}
where $x_j$'s are the zero points of the $N$-th order Hermite polynomial.
It may noted that, for the special cases with either $n=0$, $m>1,$ or $m=0$, $n>1$,
(\ref{a1}) and (\ref{a2}) reduce to the Hamiltonians of the corresponding
nonsupersymmetric spin chains. For example, in the special case with  $n=0$, $m>1$,
(\ref{a1}) and (\ref{a2}) reduce to the Hamiltonians of the
$su(m)$ bosonic HS and Polychronakos spin chain respectively.
One of the most important feature of this type of spin chains is the embedded
Yangian quantum group symmetry of their Hamiltonians,
which allows one to write down the corresponding energy spectra in closed forms
even for f\/inite number of lattice sites.
Consequently, the partition functions of such spin chains
can be expressed through Schur polynomials associated with
the irreducible representations of the Yangian quantum group
\cite{Hi95,HBM00,BBHS07}. In this context it should be noted
that, for f\/inite number of lattice sites,  in general one can not
analytically solve the Bethe ansatz equations associated with
integrable spin chains with short range interaction
and express the corresponding energy spectra
in closed form \cite{Su04,KB93,EF05}.

Since it is possible to express the spectra of the above mentioned
quantum integrable spin chains
with long range interactions in closed forms, one may naturally  enquire whether such
spectra can be generated from the energy functions associated with some classical ver\-tex mo\-dels.
Indeed, it has been found earlier that the classical partition function of a 1D vertex
model, emerging from the character of spin path conf\/igurations, is completely
equivalent to the partition function of  $su(m)$ Polychronakos
spin chain~\cite{KKN97}. Moreover, the energy levels of $su(m)$ Polychronakos
spin chain exactly match (including the degeneracy factors) with the energy
functions associated with this 1D vertex model. Similar
equivalence has been established between $su(m|n)$ supersymmetric Polychronakos
spin chain and a vertex model with suitably def\/ined
energy function~\cite{Hi2000}. It is worth noting that,
a recursion relation satisf\/ied by the partition function
of the Polychronakos spin chain plays a crucial role in establishing
the above mentioned equivalence between
quantum spin chain and  classical vertex model. However, it is not yet known
whether the partition function of HS spin chain would also satisfy similar type
of recursion relation. Therefore, some new approach is required
to search for 1D vertex model whose partition function  would be equivalent to
that of $su(m)$ HS  spin chain or its supersymmetric extension.
The purpose of the present article is to develop such a general formalism,
which can be used to f\/ind out the vertex models associated with a class of
Yangian invariant spin chains including the HS model.

\looseness=-1
The arrangement of this article is as follows. In Section~\ref{section2} we brief\/ly describe
border strips and super Schur polynomials associated with the irreducible
representations of  $Y(sl_{(m|n)})$ Yangian algebra. In Section~\ref{section3}, we def\/ine
a class of $su(m|n)$ HS like spin chains
by assuming that their partition functions
can be written in a particular form in terms of the super Schur polynomials.
Then, by using some properties of the super Schur polynomials,
 we show that the partition functions of this class of
spin chains are equivalent to the partition functions of
a class of 1D vertex models with appropriately def\/ined energy function.
Since $su(m|n)$ HS and Polychronakos spin chains belong to the above mentioned
class of Yangian invariant spin chains, we easily obtain the corresponding
vertex models
through this approach. In Section~\ref{section3}, we also establish a
boson-fermion duality relation for the partition functions
of supersymmetric HS like spin chains by using their correspondence with 1D
vertex models. In Section~\ref{section4},  we make some concluding remarks.

\section{Border strips and super Schur polynomials}\label{section2}

\begin{figure}[h!]
\centerline{\includegraphics{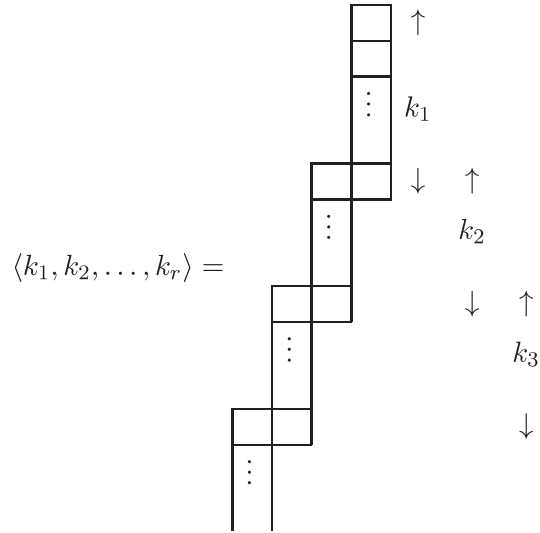}}
\caption{Shape of the border strip $\l k_1, k_2, \dots , k_r \r$.}\label{Fig1}
\end{figure}

Since in our analysis we wish to consider some HS like spin chains
with embedded $Y(sl_{(m|n)})$ symmetry, let us f\/irst brief\/ly
discuss about `border strips' which represent a class of irreducible
representations of the
$Y(sl_{(m|n)})$ algebra and span the Fock spaces of Yangian invariant
spin systems \cite{Hi95, HBM00, BBHS07, KKN97}.
For a spin system with $N$ number of lattice sites,
a border strip is uniquely characterized by
a set of positive integers $k_1, k_2, \dots, k_r$,
where $\sum_{i=1}^r k_i=N$, and $r$ is an integer which can take any value
from $1$ to $N$. Thus the vector ${\bf k} \equiv \{ k_1,\dots,k_r\}$
belongs to the set $\mathcal{P}_N$ of ordered partitions of $N$.
Let us denote the border strip corresponding
to the vector ${\bf k}$ by $\l k_1,k_2,\dots,k_r \r$, which is drawn in Fig.~\ref{Fig1}.

The super Schur polynomial corresponding to this border strip is denoted by
$S_{\l k_1, k_2, \dots, k_r \r}(x,y)$,
where $x\equiv \{x_1,\dots,x_m\}$ represents the bosonic variables
and $y\equiv \{y_1,\dots,y_n\}$ represents the fermionic variables.
For the purpose of def\/ining such super Schur polynomial,
we set $\mathbf{B}^{(m|n)} = \mathbf{B}^{(m)}_+ \cup \mathbf{B}^{(n)}_-$, where
\begin{gather}
\mathbf{B}^{(m)}_+= \{ 1,2, \dots, m \}  , \qquad
\mathbf{B}^{(n)}_- = \{ m+1,m+2, \dots, m+n \} .
\label{b1}
\end{gather}
The Young tableaux $T$ is obtained by f\/illing the numbers
$1, 2, \dots, m+n$ in a given border strip $\l k_1,k_2,\dots,k_r \r$  by the
rules:
\begin{itemize}\itemsep=0pt
\item Entries in each row are increasing, allowing the repetition of elements
in $\{i | i \in \mathbf{B}^{(m)}_+ \}$, but not permitting the repetition
of elements in $\{i | i \in \mathbf{B}^{(n)}_- \}$,
\item Entries in each column are increasing, allowing the repetition of
elements in $\{i | i \in \mathbf{B}^{(n)}_- \}$, but not permitting the
repetition of elements in $\{i | i \in \mathbf{B}^{(m)}_+ \}$.
\end{itemize}

Let $\mathcal{G}^{(m|n)}_{\l k_1,k_2,\dots,k_r \r}$ be the set of all
 Young tableaux which are obtained by f\/illing up
the border strip $\l k_1,k_2,\dots,k_r \r$ through the
above mentioned rules.
The super Schur polynomial corresponding to
$\l k_1,k_2,\dots,k_r \r$ is then def\/ined as
\begin{gather}
S_{\l k_1, k_2, \dots, k_r \r}(x,y)
= \sum_{T   \in  \mathcal{G}^{(m|n)}_{\l k_1,k_2,\dots,k_r \r}} e^{\wt(T)}  .
\label{b2}
\end{gather}
Here the weight $\wt(T)$ of the Young tableaux $T$ is given by
\begin{gather}
\wt(T) = \sum_{\alpha=1}^{m+n} \alpha(T)   \epsilon_\alpha   ,
\label{weight}
\end{gather}
where $\alpha(T)$ denotes the multiplicity of the number $\alpha$
in the Young Tableaux $T$, and we use
the notations
\begin{equation}
x_\alpha  \equiv e^{\epsilon_\alpha} \quad \text{for \ \ $\alpha \in
\mathbf{B}^{(m)}_+$}  , \qquad
y_{\alpha -m}  \equiv e^{\epsilon_{\alpha}} \quad
\text{for \ \ $\alpha \in \mathbf{B}^{(n)}_-$}.
\label{b2a}
\end{equation}
The dimensionality of the irreducible representation associated with the border
strip $\l k_1,k_2, \dots$, $k_r \r$ can be found by setting $x=1$, $y=1$ in the
corresponding Schur polynomial $S_{\l k_1, k_2, \dots, k_r \r}(x,y)$.
Hence, by using equation~(\ref{b2}), we obtain the dimension of irreducible
representation as
\begin{gather*}
S_{ \l k_1,k_2,\dots,k_r \r}(x,y) \rvert_{x=1,y=1}  =
{\mathcal N}^{(m|n)}_{\l k_1,k_2,\dots,k_r \r }   , 
\end{gather*}
where ${\mathcal N}^{(m|n)}_{\l k_1,k_2,\dots,k_r \r }$ denotes the number of
all
allowed tableaux corresponding to the border strip $\l k_1,k_2,\dots,k_r \r$.
For example, in the case of the $su(2|1)$ spin chain, it is possible to
construct the following tableaux corresponding to the border strip $\l 2,1\r$:

\vspace{1mm}

\centerline{\includegraphics{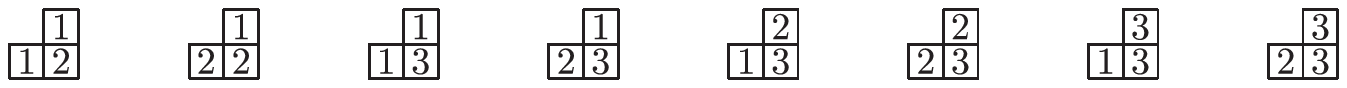}}
\vspace{1mm}

\noindent
which gives ${\mathcal N}^{(2|1)}_{ \l 2,1\r}=8$.

It is worth noting that
 the border strips can equivalently be described by
`motifs' \cite{HHBP92, Hal93},
 which for a spin chain with  $N$ number of lattice sites is given by a sequence
of $N-1$ number
of 0's and 1's, $\de = (\de_1, \de_2, \dots , \de_{N-1})$ with $\de_j \in \{
0, 1\}$. In fact, there exists a one-to-one map from a border strip to a motif
as
\begin{gather}
 \l k_1, k_2, \dots,k_r \r \ \Longrightarrow  \ \de = (\,\underbrace{1,
\dots,1}_{k_1-1}\, , 0, \underbrace{1,\dots,1}_{k_2-1},0, \dots \dots,0,
\underbrace{1,\dots,1}_{k_r-1}\,)   .
\label{b4}
\end{gather}
For example, the border strip $\l 2,3,1 \r$ is mapped to the motif $(10110)$.
Let us now def\/ine the partial sums corresponding to the
border strip $\l k_1, k_2, \dots,k_r \r$ as
\begin{gather}
\kappa_i = \sum_{l=1}^i k_l    ,
\label{b5}
\end{gather}
where $i\in \{1,2, \dots , r-1 \}$.
By using equation~(\ref{b4}) it is easy to see that,
the elements of  motif~$\de$ satisfy the following rule: $\de_j = 0$
if $j$ coincides with one of the partial sums $\kappa_i$, and $\de_j = 1$
otherwise.
The inverse mapping from a motif to a border strip can be obtained by
reading a~motif $\delta =(\delta_1, \delta_2, \dots , \delta_{N-1} )$
from the left, and adding a box under (resp. left) the box when
$\delta_j=1$ (resp.\ $\delta_j=0$) is encountered.

 \section{Equivalence between HS like spin chains\\ and some vertex models}\label{section3}

 \subsection{Def\/inition of HS like spin chains and related vertex models}\label{section3.1}

Let us assume that their exists a class of $Y(sl_{(m|n)})$
Yangian invariant HS like spin chains with $N$ number of lattice sites, for
which the
complete set of energy levels associated with  border strips like $\l k_1, k_2,
\dots,k_r \r$
(for all possible $\mathbf{k}\in \mathcal{P}_N$)
can be written in the form
\begin{gather}
E_{\l k_1,k_2,\dots,k_r \r}=\sum_{j=1}^{r-1} \mathcal{E}_N (\kappa_j)   ,
\label{c1}
\end{gather}
where $\mathcal{E}_N (j)$ is an arbitrary function
 of the discrete variable $j \in \{ 1,2, \dots , N-1 \}$.
Since any Yangian invariant spin chain must have the
same energy eigenvalue corresponding to all eigenfunctions
which form an irreducible representation of the $Y(sl_{(m|n)})$ algebra,
 the multiplicity of the eigenvalue $E_{ \l k_1,k_2,\dots,k_r \r}$ in
equation~(\ref{c1})
would be given by ${\mathcal N}^{(m|n)}_{\l k_1,k_2,\dots,k_r \r }$.
Therefore, the partition function of such a
spin chain can be expressed as
\begin{gather}
Z^{(m|n)}_N(q)= \sum_{\mathbf {k}\in P_N} q^{\sum\limits_{j=1}^{r-1} \mathcal{E}_N
(\kappa_j)}
\mathcal{N}^{(m|n)}_{ \l k_1,k_2,\dots,k_r \r}   ,
\label{c2}
\end{gather}
where $q=e^{-\frac{1}{kT}}$. We also def\/ine a `generalized partition function'
for this spin chain as
\begin{gather}
Z^{(m|n)}_N(q;x,y)=
 \sum_{\mathbf{k} \in P_N} q^{\sum\limits_{j=1}^{r-1} \mathcal{E}_N (\kappa_j)}
S_{ \l k_1,k_2,\dots,k_r \r}(x,y)   ,
\label{c3}
\end{gather}
which reduces to the usual partition function (\ref{c2}) in the special case
$x=1$, $y=1$.
It is worth noting that, all eigenvalues of the Hamiltonians (\ref{a1}) and
(\ref{a2})
can be written in the form (\ref{c1}) with
$\mathcal{E}_N(j)= j(N-j)$ and $\mathcal{E}_N(j) = j$
respectively~\cite{HBM00, BBHS07}.
Hence, both of the $su(m|n)$ supersymmetric HS  and Polychronakos spin chains
 belong to the above def\/ined class of Yangian invariant spin systems. Furthermore,
the results of~\cite{BFGR10} imply that the energy levels of a hyperbolic version of
the HS spin chain, known as Frahm--Inozemstsev chain~\cite{FI94}, can also be expressed
in the form (\ref{c1}) at least in the nonsupersymmetric limit. We shall discuss
about this spin chain in more detail in the Subsection~\ref{section3.3}.  So there exist quite a
few quantum spin chains with long range interaction, whose energy levels can  be
expressed through equation~(\ref{c1}).
However, as far as we know,  there exists no general method for constructing the
Hamiltonian of a quantum spin chain in terms of the generators of $su(m|n)$ algebra,
such that the corresponding energy levels are expressed
through equation~(\ref{c1}) for a given functional form of $\mathcal{E}_N(j)$.

Next, we consider a class of 1D classical vertex models with
$(N+1)$ number of vertices, which are connected through $N$ number of
intermediate bonds.
Each of these bonds can take either one of the $m$ possible bosonic states or
one of the $n$ possible fermionic states.
Any possible state for such $su(m|n)$ vertex model can be represented by
a path conf\/iguration of f\/inite length given~by
\begin{gather}
 \vec{s} \equiv \{ s_1,s_2,\dots,s_N\}   ,
\label{c4}
\end{gather}
where $s_i \in \{1,2,\dots , m+n \} $ denotes the spin state
of the $i$-th bond.
Let us def\/ine an `energy function' associated with the spin path conf\/iguration
$\vec{s}$ as
\begin{gather}
E^{(m|n)}(\vec{s}) = \sum_{j=1}^{N-1} \mathcal{E}_N (j)
H^{(m|n)}(s_j,s_{j+1})  ,
\label{c5}
\end{gather}
where
\begin{gather}
H^{(m|n)}(s_j,s_{j+1}) = \left\{ \begin{array}{rl}
0 & \text{if} \  s_j < s_{j+1} \  \text{or} \  s_j= s_{j+1} \in {\bf B}^{(n)}_-,
\\
1 & \text{if} \  s_j > s_{j+1} \   \text{or} \  s_j = s_{j+1} \in {\bf B}^{(m)}_+,
\end{array}  \right.
\label{c6}
\end{gather}
and $\mathcal{E}_N (j)$ is an arbitrary function of the
discrete variable $j$, which has also appeared in equation~(\ref{c1}).
The form of energy function (\ref{c5}) implies that the extreme
left and extreme right vertices, which are connected through only one bond,
always have the zero energy. The energy of all other vertices, which are connected by two
bonds, may have a f\/inite nonzero value. For example, the energy of the $j$-th vertex is
given by
\[
E^{(m|n)}_j(s_{j-1},s_j) = \tilde{\mathcal{E}}_N (j)  H^{(m|n)}(s_{j-1},s_{j})   ,
\]
where $\tilde{\mathcal{E}}_N (j)\equiv \mathcal{E}_N (j-1)$ and
  $j \in \{2,3, \dots , N \}$.
The energy function (\ref{c5})
is obtained by summing up these nontrivial vertex energies as
\[
E^{(m|n)}(\vec{s})=\sum_{j=2}^N E^{(m|n)}_j(s_{j-1},s_j)   .
\]

The partition function for this type of inhomogeneous
$su(m|n)$ vertex model may be written~as
\begin{gather}
 V^{(m|n)}_{N}(q) =\sum_{\vec{s}} q^{E^{(m|n)}(\vec{s})}   ,
\label{c7}
\end{gather}
where the summation has been taken for all possible path conf\/igurations.
In analogy with the case of Young tableaux,
one can construct a weight function for the  spin path conf\/iguration $\vec{s}$
as
\begin{equation}
\wt(\vec{s}) = \sum_{\alpha=1}^{m+n} \alpha(\vec{s})  \epsilon_\alpha   ,
\label{weight1}
\end{equation}
where $\alpha(\vec{s})$ denotes the multiplicity of the spin component $\alpha$
in the path conf\/iguration $\vec {s}$.
By using such weight functions
and  the notations given in equation~(\ref{b2a}),
we def\/ine a generalized partition function or character
for the vertex model as
\begin{gather}
V^{(m|n)}_{N}(q;x,y) =\sum_{\vec{s}} q^{E^{(m|n)}(\vec{s})} e^{\wt(\vec{s}) }  .
\label{c8}
\end{gather}
It is evident that this expression for generalized partition function
reduces to (\ref{c7}) in the special case $x=1$, $y=1$.
Let us now propose an equality between the generalized partition function
of the HS like spin chain and that of the 1D vertex model as
\begin{gather}
Z^{(m|n)}_N(q;x,y) = V^{(m|n)}_N(q;x,y)  .
\label{c9}
\end{gather}

We would like to make some comments at present.
The equality (\ref{c9}) has been established earlier
in the case of $su(m)$ and $su(m|n)$
Polychronakos spin chains for which
$\mathcal{E}_N(\kappa_j) = \kappa_j$ \cite{KKN97, Hi2000}.
In the following Subsection~\ref{section3.2}, however,
we shall prove the relation (\ref{c9}) without assuming any particular form
of the function $\mathcal{E}_N(\kappa_j)$.
Next, in  Subsection~\ref{section3.3}, we shall express the $x=y=1$ limit of equality (\ref{c9}) in an
alternative form.
Using some results of our previous works~\cite{BBHS07,BBS08},
 such alternative form of equality~(\ref{c9})
has been guessed recently for the case of nonsupersymmetric
HS like spin chains~\cite{EFG10}.

\subsection{Proof of the equivalence relation}\label{section3.2}

At f\/irst, we notice that one can map  spin path conf\/igurations
associated with vertex models
to the motifs associated with Yangian invariant spin chains
by using some well def\/ined rules~\cite{KKN97,Hi2000, EFG10}.
For our purpose, we take a slightly modif\/ied version of this mapping,
 which maps the spin path conf\/iguration $\vec{s}= \{ s_1,s_2,\dots,s_N \}$
 to the motif $\delta=(\delta_1, \delta_2, \dots,\delta_{N-1})$
 by using the rules
\begin{gather}
 (i) \ \ \text{if}  \ \ s_i < s_{i+1}  \ \ \text{or} \ \  s_i=s_{i+1} \in
\mathbf{B}^{(n)}_-
,\quad  \text{then} \ \ \delta_i= 1 , \nonumber\\
  (ii) \ \ \text{if}  \ \ s_i > s_{i+1} \ \ \text{or} \ \  s_i=s_{i+1} \in
\mathbf{B}^{(m)}_+,
\quad \text{then} \ \ \delta_i = 0   .
\label{c10}
\end{gather}
Note that the above mentioned mapping from a spin conf\/iguration
to a motif is not one-to-one. For example, in the particular case
of $su(2|0)$ model with $N=3$, both of the spin conf\/igurations
$\{112 \}$ and $\{212 \}$ are mapped to the same motif $(01)$.
 Now we consider two consecutive
lemmas which will be instrumental in proving the relation~(\ref{c9}).

\begin{lemma} \label{lemma1}
Let us assume that
a spin configuration $\vec{s} \equiv\{ s_1,s_2,\dots,s_N\}$
is mapped to the  motif $\delta \equiv (\delta_1,\delta_2,\dots ,
\delta_{N-1})$
by using the rules given in \eqref{c10}, and  this motif $\delta$ is
subsequently
mapped to the border strip $\l k_1,k_2,\dots,k_r \r$ by using
the inverse of the mapping defined in equation~\eqref{b4}. Then the
sequence of numbers $\{ s_1, s_2, \dots, s_N \}$ would generate an allowed
 tableaux for the border strip
$ \l k_1,k_2,\dots,k_r \r$, where the spin $s_i$ is put in the $i$-th box
starting from
the uppermost right side of the border strip.
\end{lemma}

\begin{proof}
 We put the spin $s_i$ in the $i$-th box and
spin $s_{i+1}$ in the $(i+1)$-th box starting from
the uppermost right side of the border strip.
The relative positions of these boxes can be either

\vspace{1mm}

\centerline{\includegraphics[scale=0.95]{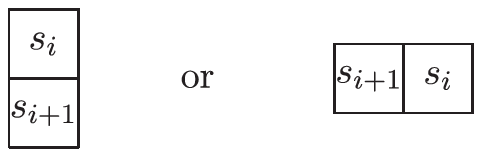}}

\vspace{1mm}


Let us f\/irst assume that either $s_i < s_{i+1}$ or $s_i=s_{i+1} \in
{\bf B}^{(n)}_-$,
which leads to $\delta_i=1$ by applying rule (i) in equation~(\ref{c10}).
According to the inverse of the mapping def\/ined in equation~(\ref{b4}),
 the $(i+1)$-th box should be placed below the $i$-th box if $\delta_i=1$:

\vspace{1mm}
\centerline{\includegraphics[scale=0.95]{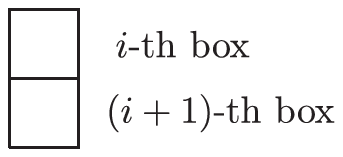}}
\vspace{1mm}

%

\noindent
Therefore, for either $s_i < s_{i+1}$ or $s_i=s_{i+1} \in {\bf B}^{(n)}_-$,
we get a part of tableaux where the $i$-th box and $(i+1)$-th box
are positioned as

\centerline{\includegraphics[scale=0.95]{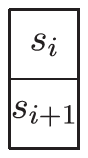}}

\noindent
It should be observed that,
 according to the rules of constructing a Young tableaux  which have been
mentioned
just after equation~(\ref{b1}),  the above f\/igure represents an allowed conf\/iguration
inside a border strip.

Next, we examine the remaining possibility, i.e.\ either
$s_i > s_{i+1}$ or $s_i = s_{i+1} \in {\bf B}^{(m)}_+ $,
which yields $\delta_i=0$ by applying rule (ii) in equation~(\ref{c10}).
According to the inverse of the mapping def\/ined in equation~(\ref{b4}),
 the $(i+1)$-th box should be placed on the left side of the $i$-th box if
$\delta_i=0$:

\vspace{1mm}
\centerline{\includegraphics[scale=0.95]{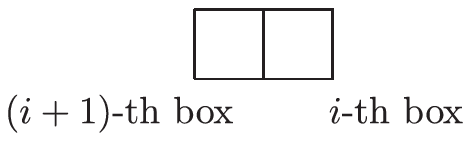}}
\vspace{1mm}


\noindent
Thus for either $s_i > s_{i+1}$ or $s_i = s_{i+1} \in {\bf B}^{(m)}_+ $,
we get a part of tableaux where the $i$-th box and $(i+1)$-th box
are positioned as

\vspace{1mm}
\centerline{\includegraphics[scale=0.95]{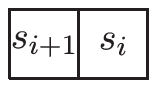}}

\vspace{1mm}


\noindent
Again, it may be noted that, according to the rules of constructing a Young
tableaux
 which have been mentioned just after equation~(\ref{b1}),
the above f\/igure represents an allowed conf\/iguration inside a border strip.
Hence we f\/ind that, for all possible choice of $s_i$ and $s_{i+1}$, the $i$-th
and $(i+1)$-th
box form a part of an allowed tableaux. Applying the aforesaid prescription
repeatedly,
we can easily show that the spin conf\/iguration $\{ s_1, s_2,\dots,s_N\}$
generates an allowed
tableaux for the associated border strip $\l k_1,k_2,\dots,k_r \r$.
\end{proof}

 It is evident that, due to Lemma~\ref{lemma1}, there exists a one-to-one mapping
from the set of all spin path conf\/igurations of the form (\ref{c4})
to the set of  Young tableaux which are constructed by f\/illing up
all possible border strips like \mbox{$\l k_1,k_2,\dots,k_r \r$}, where
$\mathbf{k}\in \mathcal{P}_N$.
It is trivial to construct the inverse of this one-to-one mapping,
which maps a Young tableaux to a spin path conf\/iguration.
By comparing equations~(\ref{weight}) and (\ref{weight1}),
we f\/ind that the weight
of a Young tableaux $T$ coincides with the weight of the corresponding
spin path conf\/iguration  $\vec{s}$:
\begin{gather}
\wt(T) = \wt(\vec{s})  .
\label{c11}
\end{gather}
Let $\mathcal{F}^{(m|n)}_{\l k_1,k_2,\dots,k_r \r}$ be the set of all
path conf\/igurations
which are mapped to a given  border strip $\l k_1,k_2,\dots,k_r \r$ by successively
applying the mapping def\/ined in equation~(\ref{c10}) and the inverse
of the mapping def\/ined in equation~(\ref{b4}). Due to Lemma~\ref{lemma1},  there exists a
one-to-one correspondence between the sets $\mathcal{F}^{(m|n)}_{\l
k_1,k_2,\dots,k_r \r}$
and $\mathcal{G}^{(m|n)}_{\l k_1,k_2,\dots,k_r \r}$. Consequently,
by using equations~(\ref{b2}) and (\ref{c11}), one can express the
super Schur polynomial corresponding to the border strip $\l k_1,k_2,\dots,k_r
\r$ as
\begin{gather}
S_{\l k_1,k_2,\dots,k_r \r} (x,y) = \sum_{   \vec{s}
  \in   \mathcal{F}^{(m|n)}_{\l k_1,k_2,\dots,k_r \r} }   e^{\wt(\vec{s}) }   .
\label{c12}
\end{gather}
Now, we consider the second lemma:

\begin{lemma}\label{lemma2}
If the spin configuration $\vec{s}\equiv
\{s_1,s_2,\dots,s_N\}$ is mapped
to the border strip $\l k_1,k_2,\dots,$ $k_N \r$ by successively applying the
mapping defined in equation~\eqref{c10} and the inverse
of the mapping defined in equation~\eqref{b4},
then the energy function \eqref{c5} associated with $\vec{s}$ can be expressed~as
\begin{gather}
E^{(m|n)}(\vec{s}) =\sum_{l=1}^{r-1} \mathcal{E}_N (\kappa_l).
\label{c13}
\end{gather}
\end{lemma}

\begin{proof}
By comparing equation~(\ref{c6}) with equation~(\ref{c10}),
one can write $H^{(m|n)}(s_j,s_{j+1})$ as
\[
H^{(m|n)}(s_j,s_{j+1} )=1-\delta_j.
\]
Substituting this expression of $H^{(m|n)}(s_j,s_{j+1})$ to
equation~(\ref{c5}), we obtain
\begin{gather}
E^{(m|n)}(\vec{s})
=\sum_{j=1}^{N-1} \mathcal{E}_N (j) (1-\delta_j)   .
\label{c14}
\end{gather}
Thus, only the zeros of the motif $(\delta_1,\delta_2,\dots,\delta_{N-1})$
would
contribute to $E^{(m|n)}(\vec{s})$. We have already mentioned that in Section~\ref{section2}
that,
 the positions of the zeros of a motif coincide with the partial sums (\ref{b5})
associated with the corresponding border strip
$\l k_1,k_2,\dots,k_r \r$.  Therefore, we can write $(1-\delta_j)$ in terms of
Kronecker $\delta$-function as
\[
1-\delta_j = \sum_{l=1}^{r-1} \delta_{j, \kappa_l}   .
\]
Substituting this expression of $(1-\delta_j)$ to equation~(\ref{c14}),
and summing over the index $j$
by using Kronecker $\delta$-function, we easily get equation~(\ref{c13}).
The form of equation~(\ref{c13}) ensures that,
for all spin path conf\/igurations $\{s_1,s_2,\dots,s_N\}\in
\mathcal{F}^{(m|n)}_{\l k_1,k_2,\dots,k_r \r}$,
we get the same value of the energy function.
\end{proof}

Now we are in a position to prove the main proposition of this paper, which is
given in equation~(\ref{c9}). To this end we note that, as a consequence of Lemma~\ref{lemma1},
equation~(\ref{c8}) can be written by rearranging its summation variables as
\begin{gather}
V^{(m|n)}_{N}(q;x,y) = \sum_{ \mathbf {k}\in P_N}  \sum_{\vec{s}
\in \mathcal{F}^{(m|n)}_{\l k_1,k_2,\dots,k_r \r}}  q^{E^{(m|n)}(\vec{s})}
e^{ \wt(\vec{s}) }   .
\label{c15}
\end{gather}
Substituting the expression of $E^{(m|n)}(\vec{s})$ given in equation~(\ref{c13})
to equation~(\ref{c15}), we f\/ind that
 \begin{gather}
V^{(m|n)}_{N}(q;x,y) = \sum_{ \mathbf {k}\in P_N}  q^{\sum\limits_{l=1}^{r-1}
\mathcal{E}_N (\kappa_l)}
 \sum_{\vec{s}
\in \mathcal{F}^{(m|n)}_{\l k_1,k_2,\dots,k_r \r}}   e^{\wt(\vec{s})}   .
\label{c16}
\end{gather}
Finally, by using the relation (\ref{c12}),  we can express equation~(\ref{c16}) as
 \begin{gather}
V^{(m|n)}_{N}(q;x,y) = \sum_{ \mathbf {k}\in P_N}  q^{\sum\limits_{l=1}^{r-1}
\mathcal{E}_N (\kappa_l)}
 S_{\l k_1,k_2,\dots,k_r \r} (x,y)   .
\label{c17}
\end{gather}
Since the r.h.s. of equation~(\ref{c17}) coincide with that of equation~(\ref{c3}),
 we readily obtain a proof of equation~(\ref{c9}). By putting $x=y=1$ in
equation~(\ref{c9}),
we also get
\begin{gather}
Z^{(m|n)}_N(q) = V^{(m|n)}_N(q)   ,
\label{c18}
\end{gather}
which shows that the partition functions of all Yangian invariant
HS like spin chains, with energy levels given by equation~(\ref{c1}),
exactly coincide with the partition functions of 1D vertex models with
energy functions given by equation~(\ref{c5}). Moreover, since
$Z^{(m|n)}_N(q)$ or $ V^{(m|n)}_N(q)$ can be expressed as a power series of $q$,
where the powers represent all possible energy levels or energy functions,
it is clear from the equality (\ref{c18}) that the energy levels of the HS like
spin chains exactly match with the energy functions of the corresponding
vertex models.

We have already mentioned that, all eigenvalues of the Hamiltonians (\ref{a1})
and (\ref{a2})
can be written in the form (\ref{c1}), with
$\mathcal{E}_N(j)= j(N-j)$ and $\mathcal{E}_N(j) = j$
respectively. Hence, due to equation~(\ref{c18}) it follows that, the partition
functions of
$su(m|n)$ supersymmetric HS and Polychronakos spin chains
would be equivalent to those of $su(m|n)$ vertex models
with energy functions given by (\ref{c5}),
where $\mathcal{E}_N(j)= j(N-j)$ and $\mathcal{E}_N(j) = j$ respectively.
It should be noted that,  since  the relations
(\ref{c9}) and (\ref{c18}) have been derived by assuming that $m$ and $n$ are
arbitrary non-negative integers (excluding the nonphysical case $m=n=0$),
these relations are also valid for the nonsupersymmetric cases like
$n=0$, $m>1$ or $m=0$, $n>1$. Therefore, by considering
equation~(\ref{c18}) for such special cases,
one can also establish an equivalence between the partition functions of
nonsupersymmetric HS like spin chains and those of the corresponding 1D vertex
models.

 \subsection{An alternative form of the equivalence relation}\label{section3.3}

Let us now try to express the remarkable relation (\ref{c18}) in an alternative
form. It is well known that the `freezing trick' provides a  powerful method
of calculating the partition functions of
quantum integrable spin chains with long range interaction \cite{Po94}.
By using this freezing trick,
the partition functions of supersymmetric and nonsupersymmetric spin systems
associated with the $A_N$ type of root system can often be expressed in the form
\cite{FG05,BMN06,BBS08, BFGR08}
\begin{gather}
\widetilde{Z}^{(m|n)}(q)  = \sum _{\mathbf{k} \in   \mathcal{P}_N}
\prod_{i=1}^r d^{(m|n)}(k_i)   \cdot
q^{\sum \limits^{r-1}_{j=1}\mathcal{E}(\kappa_j) }  \prod_{i=1}^{N-r}
\big(1-q^{\mathcal{E}(\kappa'_i) }\big)   ,
\label{c19}
\end{gather}
where $\kappa'_1, \kappa'_2, \dots, \kappa'_{N-r}$ are the complements of the
partial sums def\/ined in equation~(\ref{b5}), i.e.
$\kappa'_1, \kappa'_2, \dots, \kappa'_{N-r} \equiv
\{1,2, \dots,N-1\}-\{\kappa_1, \kappa_2,\dots ,\kappa_{r-1}\}$, and
$d^{(m|n)}(k_i)$ is given by
\begin{gather*}
 d^{(m|n)}(k_i)  = \left\{ \begin{array}{ll}
 {}^m C_{k_i}  , & \text{if} \ n=0,\  m>1,   \\
{}^{n+k_i-1} C_{k_i}   , & \text{if} \ m=0, \  n>1   ,\\
 \sum\limits_{j=0}^{{\rm min} (m,k_i)} {}^m C_j ~{}^{k_i-j+n-1}
C_{k_i-j}  , & \text{if} \  m, n>0   ,
\end{array}  \right.
\end{gather*}
with ${}^p C_l \equiv \frac{p!}{l  (p-l)!} $ for $l\leq p$ and
${}^p C_l \equiv 0$ for $l>p$.
It can be shown that, for any possible choice of the function $\mathcal{E}(j)$,
 the partition functions given by equations~(\ref{c2}) and (\ref {c19})
are completely equivalent to each other \cite{BBHS07,BBS08}:
\begin{gather} Z^{(m|n)}(q) = \widetilde{Z}^{(m|n)}(q)   .
\label{c20a}
\end{gather}
Due to this equivalence, we can express equation~(\ref{c18}) in an
alternative form like
\begin{gather}
\widetilde{Z}^{(m|n)}(q) = V^{(m|n)}(q)   .
\label{c21}
\end{gather}
It should be noted that,
the nonsupersymmetric version of the above
relation has been stated recently (without an explicit proof) and also
used to analytically f\/ind out the level density distribution
of some HS like spin chains with large number of lattice sites \cite{EFG10}.

In this context we would like to mention that,
there exist a hyperbolic variant \cite{FI94} of
the HS spin chain, whose partition function has been calculated
 for the nonsupersymmetric case by using the freezing trick~\cite{BFGR10}.
The Hamiltonian of such hyperbolic version of
the HS spin, known as Frahm--Inozemstsev (FI) chain, can be written
for the most general $su(m|n)$ supersymmetric case as
\begin{gather*}
\mathcal {H}_{\rm FI}^{(m|n)} = \frac{1}{2} \sum_{j<k}
\frac{1 + \hat{P}^{(m|n)}_{jk}}{\sinh^2(\xi_j-\xi_k)}  ,
\end{gather*}
where the lattice sites of this chain
are given by $\xi_i=\frac{1}{2} \ln \zeta_i$,
and $\zeta_i$ is the $i$-th zero of the Laguerre polynomial $L^{\alpha -1}_N$
with $\alpha >0$. It has been found that, for
 nonsupersymmetric cases like $n=0$, $m>1$ and $m=0$, $n>1$,
partition functions of this spin chain
can be obtained in the form (\ref{c19}),
with $\mathcal{E}(j) = j(\alpha +j -1)$ \cite{BFGR10}. Hence, by using
equation~(\ref{c20a}), one can show that
 the energy levels of nonsupersymmetric FI spin chains
are expressed through equation~(\ref{c1}) with exactly same form of $\mathcal{E}(j)$.
Moreover, due to the relation  (\ref{c21}),
it is evident that the corresponding partition functions would be equivalent
to those of $su(m|0)$ and $su(0|n)$
vertex models with energy functions given by  equation~(\ref{c5}),
where $\mathcal{E}(j) = j(\alpha +j -1)$.

 \subsection{Boson-fermion duality relation}\label{section3.4}

Let us now discuss an interesting application of the above mentioned
equivalence between the partition functions of HS like spin chains and
1D vertex models. It is known that the partition  functions of the
$su(m|n)$ supersymmetric HS and Polychronakos spin chains
satisfy a remarkable duality relation under the exchange of bosonic
and fermionic spin degrees of freedom \cite{BMUW99, HBM00, BMN06,BBHS07,Hal93}.
Here, we want to show that the partition functions of all
HS like spin chains, which can be expressed
in the form (\ref{c2}) or (\ref{c19}), would satisfy this type
of boson-fermion duality relation. To this end, let us def\/ine a
mapping which maps a spin path conf\/iguration $\vec{s} \equiv \{
s_1,s_2,\dots,s_N\}$
of $su(m|n)$ vertex model to a spin path conf\/iguration
$\vec{t} \equiv \{ t_1,t_2,\dots,t_N\}$ of $su(n|m)$ vertex model~as
\begin{gather}
t_i= (m+n+1) -s_i  ,
\label{c23}
\end{gather}
where $i\in \{1,2, \dots , N\}$.  Note that (\ref{c23})
represents a one-to-one and invertible mapping from the set of all spin path
conf\/igurations of $su(m|n)$ vertex model to the set of all
spin path conf\/igurations of $su(n|m)$ vertex model. Moreover,
it may also be observed that, if $s_i \in B_+^{(m)} $
 (resp.\ $s_i \in B_-^{(n)} $) then  $t_i \in B_-^{(m)} $ (resp.\ $t_i \in
B_+^{(n)}$),
and if $s_i<s_{i+1}$ (resp.\ $s_i>s_{i+1}$) then
$t_i>t_{i+1}$ (resp. $t_i <t_{i+1}$). Hence, due to equation~(\ref{c6}),
it follows that
\[
H^{(m|n)}(s_j,s_{j+1})  =1 - H^{(n|m)}(t_j,t_{j+1})  .
\]
Substituting the above relation to equation~(\ref{c5}), we obtain
\begin{gather}
E^{(m|n)}(\vec{s}) = \Omega  - E^{(n|m)}(\vec{t})   ,
\label{c24}
\end{gather}
where $\Omega$ is given by
\begin{gather}
\Omega=\sum_{j=1}^{N-1}\mathcal{E}_N(j)   .
\label{c25}
\end{gather}
By using equations~(\ref{c7}) and (\ref{c24}), we f\/ind that there exists
a boson-fermion duality relation between the partition functions of
$su(m|n)$ and $su(n|m)$ vertex models:
\begin{gather*}
V^{(m|n)}(q) = q^\Omega ~ V^{(n|m)}\big(q^{-1}\big)   .
\end{gather*}
Due to the equality (\ref{c18}), the above equation
can be transformed to a boson-fermion duality relation
between the partition functions of $su(m|n)$ and $su(n|m)$
spin chains as
\begin{gather}
Z^{(m|n)}(q) = q^\Omega ~Z^{(n|m)}\big(q^{-1}\big)   .
\label{c27}
\end{gather}
Thus we are able to prove that, the partition functions of all
HS like spin chains, which can be expressed
in the form (\ref{c2}), would satisfy the duality relation (\ref{c27})
where $\Omega$ is obtained by using equation~(\ref{c25}).
Due to equation~(\ref{c21}) it follows that, the partition function
$\widetilde{Z}^{(m|n)}(q)$ given in equation~(\ref{c19})
would also satisfy the exactly same form of duality relation.
Since the partition functions of $su(m|n)$ HS and Polychronakos spin chains
can be written in the form (\ref{c2}) with
$\mathcal{E}_N(j)= j(N-j)$ and $\mathcal{E}_N(j)= j$ respectively, it is evident
 that these partition functions satisfy the boson-fermion duality relation
(\ref{c27}) with $\Omega = \sum_{j=1}^{N-1} j(N-j) = N(N^2-1)/6$
and  $\Omega = \sum_{j=1}^{N-1} j = N(N-1)/2$ respectively.

\section{Concluding remarks}\label{section4}

In this article we  consider a class of $Y(sl_{(m|n)})$
Yangian invariant spin chains, which contain the well known
 $su(m|n)$ supersymmetric Haldane--Shastry (HS) and Polychronakos spin chains as
special cases. Since the irreducible representations of the
$Y(sl_{(m|n)})$ algebra can be characterized through the border strips,
partition functions of this class of HS like spin chains can be
expressed in terms of the super Schur polynomials associated with the border
strips. Representing such super Schur polynomial as a summation over spin path
conf\/igurations,
we establish the key relations~(\ref{c9}) and~(\ref{c18}),
 which show that the partition functions of this class of
spin chains are equivalent to the partition functions of
a class of one-dimensional (1D) vertex models with appropriately def\/ined energy
functions. Since  $su(m|n)$ supersymmetric HS and Polychronakos spin chains belong to this
class of Yangian invariant spin chains, we easily obtain the corresponding
vertex models through this approach. Moreover,
the vertex model associated with FI spin chain can also be obtained through this
approach, at least in the nonsupersymmetric case.

In this context it is interesting to recall that,
there exists a close connection between 1D quantum integrable spin chains
with short range interaction and 2D vertex models in classical statistical
mechanics \cite{Su70, Ba71}. However, for these cases,
there exists no simple relation between the
energy levels of 1D  spin chains and the energy functions of 2D vertex models.
 On the other hand, the presently derived relation
(\ref{c18}) implies that, all energy levels of a class of HS like spin chains with
long range interaction exactly coincide (including the degeneracy
factors) with the energy functions of some
1D inhomogeneous vertex models in classical statistical mechanics.
Hence, this connection between 1D quantum spin chains and 1D classical vertex models
is much more direct in nature. By using this connection, we have derived
a boson-fermion duality relation (\ref{c27}) for the partition functions
of this class of supersymmetric HS like spin chains.
Moreover, it may be noted that, the nonsupersymmetric limit of this connection has been
used very recently to analytically f\/ind out the level density distribution
of some bosonic and fermionic HS like spin chains~\cite{EFG10}. Therefore,
it is natural to expect that the more general equivalence relation~(\ref{c18})
would also be useful to analytically f\/ind out the level density distribution
of supersymmetric HS like spin chains. Finally,
this type of connection between
HS like spin chains with long range interaction and 1D vertex models
may also be useful in calculating various correlation functions
and statistical properties of these systems.

\pdfbookmark[1]{References}{ref}
\LastPageEnding

\end{document}